\documentclass[12pt,preprint,flushrt]{aastex}

\begin{document}
\newcommand{\etal}{{\it et al.}}
\include{idl}

\title{Evidence for Orbital Decay of RX J1914.4+2456: Gravitational Radiation 
and the Nature of the X-ray Emission}
\author{Tod E. Strohmayer}
\affil{Laboratory for High Energy Astrophysics, NASA's Goddard Space Flight 
Center, Greenbelt, MD 20771; stroh@clarence.gsfc.nasa.gov}

\begin{abstract}

RX J1914.4+2456 is a candidate double-degenerate binary (AM CVn) with a 
putative 569 s orbital period. If this identification is correct, then it has 
one of the shortest binary orbital periods known, and gravitational radiation 
should drive the orbital evolution and mass transfer if the binary is 
semi-detached. Here we report the results of a coherent timing study of the 
archival ROSAT and ASCA data for RX J1914.4+2456. We performed a phase 
coherent timing analysis using all observations spanning an $\approx 4.6$ year 
period. We demonstrate that all the data can be phase connected, and we 
present evidence  that the 1.756 mHz orbital frequency is increasing at a rate 
of $8 \pm 3 \times 10^{-18}$ Hz s$^{-1}$, consistent with the expected loss 
of angular momentum from the binary system via gravitational radiation. 
In addition to providing evidence for the emission of gravitational waves,  
measurement of the orbital $\dot\nu$ constrains models for the X-ray 
emission and the nature of the secondary. If stable mass accretion drives the 
X-ray flux, then a positive $\dot\nu$ is inconsistent with a degenerate donor. 
A helium burning dwarf is compatible if indeed such systems can have periods as
short as that of RX J1914.4+2456, an open theoretical question. Our measurement
of a positive $\dot\nu$ is consistent with the unipolar induction model of Wu
et al. which does not require accretion to drive the X-ray flux. We discuss how
future timing measurements of RX J1914.4+2456 (and systems like it) with 
for example, Chandra and XMM-Newton, can provide a unique probe of the 
interaction between mass loss and gravitational radiation. We also discuss the
importance of such measurements in the context of gravitational wave detection
from space, such as is expected in the future with the LISA mission.

\end{abstract}

\keywords{Binaries: general - Stars: individual (RX J1914.4+2456) - Stars: 
white dwarfs - cataclysmic variables - X-rays: stars - X-rays: binaries}

\section{Introduction}

The evolution of highly compact binary stars is driven to a large extent by the
interplay between angular momentum loss from gravitational radiation and 
mass transfer between the components (see, for example Rappaport, Joss \&
Webbink 1982). Detailed study of the orbital evolution of such systems can 
thus provide a unique physical laboratory for the study of gravitational 
radiation and its effects on binary evolution. 

The shortest period cataclysmic variables (CVs), the AM CVn 
stars, are the most compact binaries known, with orbital periods $< 40$ 
minutes. Many may be double-degenerate systems (see Warner 1995 for a review). 
Their formation is complex, depending on unstable mass transfer leading to 
common envelope (CE) evolution, processes which are not well understood 
(see Nelemans et al. 2001a). Such systems likely begin as a pair of main 
sequence stars, each of 3 - 5 $M_{\odot}$. After perhaps two episodes of CE 
evolution a pair of white dwarfs separated by only a few $R_{\odot}$ remain. 
Gravitational radiation losses eventually bring the pair closer until the 
smaller mass star begins to fill its Roche lobe. The minimum period 
(at contact) of such a system depends on the details of the constituents, but 
it may be as short as 5-6 minutes (Tutukov \& Yungelson 1996; Hils \& Bender 
2000). 

Theoretical studies by Tutukov \& Yungelson (1996) and more recently, 
Nelemans et al. (2001a) suggest that as many as $\sim 10^8$ double-degenerate 
systems may populate the Galaxy. Because of their compact nature, these objects
are ideal targets for space based gravitational wave detection with the Laser 
Interferometer Space Antenna (LISA) mission. They are likely the progenitors of
at least some type Ia supernovae and may also represent a substantial fraction 
of supersoft X-ray sources (see Hils \& Bender 2000; Nelemans, Yungelson \& 
Portegies Zwart 2001b). Their evolution has also been proposed as a channel 
for production of millisecond radio pulsars via accretion induced collapse 
(see Savonije, de Kool, \& van den Heuvel 1986). An understanding of their 
formation, evolution and properties is therefore crucial to many areas of 
active astrophysical investigation.

The recent discovery of two double-degenerate binary candidates, 
RX J1914.4+2456 with a period of 569 s, and RX J0806+15 with a period of 
321 s, has sparked keen interest in these objects (see Israel et al. 1999;
Israel et al. 2002; Ramsay, Hakala \& Cropper 2002). RX J1914.4+2456 
(hereafter RX J1914) was discovered in the ROSAT all-sky 
survey. Motch et al. (1996) found that the X-ray flux is modulated with a 
period of $\approx 569$ s, and concluded that the source was likely an 
intermediate polar (IP), with the 569 s period reflecting the spin of
the white dwarf primary. However, subsequent ROSAT observations did not find 
any additional periods such as are commonly present in IPs. This, combined 
with the unusual 100\% X-ray modulation, led Cropper et al. (1998) to suggest 
that the source is a double-degenerate polar. Polars, systems containing an 
accreting strongly magnetized white dwarf, rotate synchronously with the 
orbital period. If this identification is correct, RX J1914 has one of 
the shortest binary orbital periods known. If the orbital period is 569 s, and 
the secondary fills its Roche lobe, it must be a helium star, and perhaps
a degenerate He white dwarf. Savonije et al. (1986) predict a 
minimum period of 636 s for a non-degenerate helium burning secondary, which,
given the theoretical uncertainties, may not be long enough to completely rule 
out a non-degenerate secondary as argued by Cropper et al. (1998). 

The X-ray spectrum of RX J1914 is soft, with a 40 eV blackbody model providing 
an acceptable fit to the ROSAT PSPC data (Motch et al. 1996; Cropper et al. 
1998). The X-ray luminosity is in the range $4\times 10^{33}$ to 
$1\times 10^{35}$ ergs s$^{-1}$ for an assumed distance of 100 to 400 pc. 
Accretion at a rate of $\sim 1\times 10^{-8} M_{\odot}$ yr$^{-1}$ could 
conceivably power the X-ray flux. Ramsay et al. (2000) detected the optical 
counterpart to RX J1914. It is modulated in the V, R, and I bands at the 569 s 
period, but the optical peak leads the X-ray peak by about 0.4 cycles. This
strengthens the orbital identification for the 569 s period. 
No other periods are seen in the optical data. More recent spectroscopy by 
Ramsay et al. (2002) reveals a line-free spectrum and no detectable 
polarization, inconsistent with its identification as a polar. 
If the source is indeed a polar, then the lack of polarization and emission 
lines is puzzling. Alternatively, Marsh \& Steeghs (2002) have recently
suggested that the primary may be non-magnetic and accretion takes place via
direct impact of the accretion stream (i.e., a double-degenerate Algol). 
An alternative not requiring accretion has been proposed by Wu et al. (2002).
They suggest the system might be a unipolar inductor, with the secondary 
moving through the magnetic field of the primary because of a small 
asynchronism between the orbit and primary spin (in analogy with the Jupiter
- Io system). This sets up an electric field which can drive currents in the
magnetosphere, providing the energy for the X-ray emission. Most recently,
Norton, Haswell \& Wynn (2002) have suggested that the system may be a 
stream-fed IP viewed nearly face-on. If this hypothesis is correct, then the
orbital period is not seen directly, and the observed X-ray periodicity is 
the beat period between the spinning white dwarf primary and the orbital 
period. 

Without any additional information it is difficult to determine which, if any,
of the current hypotheses regarding the nature of RX J1914 is correct. 
However, strong clues to the nature of the X-ray flux could be obtained if
the orbital evolution of the system can be detected.  Indeed,
for conservative mass transfer it is relatively straightforward to show that 
if the donor star is a degenerate dwarf, then stable mass transfer can only 
proceed with a widening of the orbit and a decrease in the orbital frequency 
(see Nelemans et al. 2001a; Savonije et al. 1986). Thus, a measurement of the 
rate of change of the orbital frequency can provide a way to break the model 
degeneracy outlined above.

This possibility has led us to reinvestigate the available timing data on 
RX J1914, with the goal of trying to detect or constrain orbital frequency 
changes in the system. Here we report the results of our timing analysis of the
ROSAT and ASCA observations of RX J1914 over a period of $\approx 4.6$ years. 
We find evidence for an increase in the orbital frequency of the system 
at a rate of $\approx 8 \times 10^{-18}$ Hz s$^{-1}$, which is similar to that 
expected from gravitational radiation losses if the observed X-ray period is
indeed the orbital period. 

The plan of this paper is as follows. In \S 2 we describe in 
detail our phase coherent timing study of the ROSAT and ASCA data. We show 
that all the data can be phase connected and that a positive $\dot\nu$ is 
strongly indicated. In \S 3 we discuss the implications of our findings for 
the nature of the X-ray emission from RX J1914 and show that orbital decay 
argues against accretion as the source of the X-ray flux unless the donor is 
non-degenerate. We conclude in \S4 with a summary of our principal findings and
plans for continued precision timing of such objects, and the synergy of such 
measurements with future space-based gravitational wave detection.

\section{Data Extraction and Analysis}

A total of $\approx 100$ ksec of observing time on RX J1914 was obtained with 
ROSAT and ASCA over the period from September, 1993 to April, 1998. Of the 
six observations, one was with the ROSAT Position Sensitive Proportional 
Counter (PSPC), four were ROSAT High Resolution Imager (HRI) pointings, and the
most recent observation was with the ASCA Solid-State Imaging Spectrometer 
(SIS). These data are now all in the public HEASARC archive. A number of 
studies of some or all of these data have been presented in the literature. 
Motch et al. (1996) discovered the 569 s modulation. Cropper et al. (1998) 
and Ramsay et al. (2000) have also reported timing results for RX J1914. 
Ramsay et al. (2000) used all the available ROSAT data, in addition to the
single ASCA pointing and found that they could phase-connect all the data to 
within $\approx 0.02$ cycles using a constant frequency model.

Table 1 gives a log of the observations used in our study. We used the
HEASARC FTOOLS data analysis package (i.e., XSELECT) to extract and 
analyse the data.  We began by producing images and extracting source 
events.  We extracted events from circular regions around the point source
centroid using radii consistent with the relevant point spread functions of 
each instrument. In all cases the point source was easily identified and 
there were no source confusion problems. For example, Figure 1 shows the HRI 
image obtained from the April 30, 1996 observation (26 ksec exposure). 

We next applied barycentric corrections to the event times for each 
observation. We used the standard mission ftools in conjunction with the 
ROSAT and ASCA orbital and JPL (DE200) solar system ephemerides 
(Standish et al. 1992). For the source position we used the coordinates,
($\alpha = 19^h 14^m 26.1^s$, $\delta = 24^{\circ} 56' 43.6''$ :J2000),
obtained by Ramsay et al. (2002) from their study of the optical-infrared 
counterpart to RX J1914. The ROSAT tool {\it abc}, which computes the X-ray 
event times at the solar system barycenter, also corrects for the spacecraft 
clock drift. With the drift removed, the ROSAT clock is typically 
stable to a few 10s of milliseconds. Even a clock error of a 100 milliseconds 
between observing epochs corresponds to only a $1.8\times 10^{-4}$ cycle 
error for RX J1914. Thus, the ROSAT times are more than precise enough for 
our purposes. Hirayama et al. (1996) estimate the ASCA absolute timing 
precision to be on the order of 2 ms. Moreover, Saito et al. (1997) found the 
stability of the ASCA clock frequency on a timescale $T$ to be, 
$\delta f / f < 3 \times 10^{-8} (T/10^4 \; {\rm s})^{-1}$. This level of drift
would produce a fractional phase shift during the ASCA observation of only 
$< 1\times 10^{-6}$. These numbers indicate that the ASCA timing capability is
at least as good as the expected ROSAT precision, and is more than sufficient
for our timing study of RX J1914. Our extraction of source events from the 
six observations resulted in a total of 4315 photons.

\subsection{Coherent Timing Methods}

We performed our coherent timing studies using the $Z_n^2$ 
statistic (Buccheri 1983; see also Strohmayer \& Markwardt 2002 for an 
example of the use of this statistic in a similar context). In our study we 
performed model fitting in two complementary ways; a total power method, and 
a phase fitting method similar to those used in pulsar timing studies. Although
both methods use the same data, the total power method does not directly 
utilize the phase information. We perform both analyses in order to provide
a means of double checking our results and to have a level of ``redundancy,'' 
however, the phase timing method uses all the available information 
(both magnitude and phase) and we generally will describe our quantitative 
results in terms of this technique. 

For the total power method we evaluate;
\begin{equation}
Z_n^2 = \frac{2}{N} \sum_{k=1}^n \left [ \left ( \sum_{j=1}^N \cos k \phi_j 
 \right )^2 + \left ( \sum_{j=1}^N \sin k \phi_j \right )^2 \right ] \; ,
\end{equation}
where $\phi_j = 2\pi \int_0^{t_j} \nu (t') dt$, $\nu (t')$ is the 
frequency evolution model, $t_j$ are the observed X-ray event times, $N$ is
the total number of X-ray events, and $n$ is the number of harmonics 
included in the sum. With this method we vary the timing model 
parameters in order to maximize the total $Z_n^2$ power. 

For our phase timing analysis we begin by defining the complex vector
\begin{equation}
C_n = \sum_{k=1}^n \left ( \sum_{j=1}^N \cos k\phi_j  + i \ \sum_{j=1}^N 
\sin k\phi_j \right ) \; . 
\end{equation}
The phase angle, $\psi$ is then defined as 
\begin{equation}
\psi = \tan^{-1} \left [ \frac{\sum_{k=1}^n \sum_{j=1}^N \sin k\phi_j}
{\sum_{k=1}^n \sum_{j=1}^N \cos k\phi_j} \right ] \; .
\end{equation}
To perform a phase timing fit we break up the X-ray event times into a set 
of $M$ bins and compute the phase angle $\psi_l$ for each bin. The phases are 
simply given by;
\begin{equation}
\psi_l = \tan^{-1} \left [ \frac{\sum_{k=1}^n \sum_{m} \sin k\phi_m}
{\sum_{k=1}^n \sum_{m} \cos k\phi_m} \right ] \; ,
\end{equation}
where the index $m$ runs over all events in bin $l$. We then compute
\begin{equation}
\chi^2 = \sum_{l=1}^M \left ( \psi_l - \psi_{avg} \right )^2 / 
\sigma_{\psi_l}^2 \; , 
\end{equation}
where $\psi_{avg}$ is the average phase angle computed 
from all $M$ bins. To find the best fitting model, we vary the timing 
parameters and search for those which yield the minimum $\chi^2$. 
For a coherent signal the error $\sigma_{\psi_l}$ in the phase angle 
is given simply by $1/\sqrt(Z_n^2)$. 

In modelling the ROSAT and ASCA event times we use a two parameter frequency 
model; $\nu (t) = \nu_0 + \dot\nu (t - t_0)$, where $\nu_0$, $\dot\nu$ and 
$t_0$ are the orbital frequency at $t_0$, the orbital frequency derivative, 
and the reference epoch, respectively. With this model the phase advance due 
to $\dot\nu$ has the well known quadratic time dependence; 
$\Delta\phi = \frac{1}{2} \dot\nu (t-t_0)^2$. 

\subsection{Theoretical Expectations}

For a detached binary with a circular orbit the rate of change of the orbital 
frequency due to gravitational radiation is (see for example, Evans, Iben \&
Smarr 1987; Taylor \& Weisberg 1989)
\begin{equation} 
\dot\nu_{gr} = 1.64 \times 10^{-17} \ \left ( \frac{\nu}{10^{-3}\; 
{\rm Hz}} \right )^5 \; \left ( \frac{\mu}{M_{\odot}} \right ) \; 
\left ( \frac{a}{10^{10} \; {\rm cm}} \right )^2 \;\; {\rm Hz} 
\ {\rm s}^{-1} \; ,
\end{equation}
where, $\nu$, $\mu$, and $a$ are the orbital frequency, reduced mass and 
orbital separation of the components, respectively. Given the 569 s 
orbital period of RX J1914, and likely scenarios for the binary components,
a total system mass of $\lesssim 1 M_{\odot}$ and reduced mass of 
$\mu \approx 0.05 M_{\odot}$ are likely (see Cropper et al. 1998; 
Wu et al. 2002). With these numbers, Kepler's law gives an estimate of the 
orbital separation $a \approx 1 \times 10^{10}$ cm. This leads to an estimate 
of $\dot\nu_{gr} \approx  1 \times 10^{-17}$. A $\dot\nu$ of this size 
would produce a total phase advance of $\approx 0.1$ cycles over the four year
timespan of the data. We note again that this is much larger than any 
anticipated errors associated with drift of the ROSAT or ASCA clocks. 

As noted earlier orbital evolution involves the interplay between
mass transfer and angular momentum loss to gravitational radiation. Indeed,
for conservative mass transfer it is relatively straightforward to show that if
the donor star is a degenerate dwarf, then stable mass transfer can only 
proceed with a widening of the orbit and a decrease in 
the orbital frequency (see Nelemans et al. 2001a; Savonije et al. 1986). If 
there is no mass transfer then one would expect the orbital frequency to 
increase (positive $\dot\nu$). In either case the {\it magnitude} of the 
orbital frequency evolution is set by $\dot\nu_{gr}$ above. 

\subsection{Results}

We began our analysis by conducting a total $Z_n^2$ power search using the
constant frequency model ($\dot\nu = 0$). We searched in a frequency range 
around the known 0.00659 day (1.756 mHz) period reported by Ramsay et al. 
(2000), and we sampled with a resolution substantially finer than the 
anticipated frequency resolution of $1/T$, where $T$ is the total timespan of 
the data (about 4.6 years). We found a best frequency of $1.7562467 
\times 10^{-3} \pm 2 \times 10^{-10}$ Hz, which is consistent with the 
constant frequency found by Ramsay et al. (2000). 

By varying $n$ we found that most of the signal was in the fundamental and
first harmonic, so for the remainder of our analysis we fixed $n=2$. 
Figure 2 shows the $Z_2^2$ power spectrum in the vicinity of our best period. 
Note the multiple side-lobe peaks which result from the sparse sampling (i.e., 
time gaps). Since the sampling is so sparse it is crucial to 
investigate the range of both $\nu$ and $\dot\nu$ which could conceivably 
fit the data. 
In order to bound the phase space in which to make a coherent search with 
all the data we first computed a frequency history by analysing each 
observation individually. We then fit the individual time - frequency 
measurements to a linear frequency evolution model. 
We found that the frequency history is well fit by a constant frequency 
($\dot\nu = 0$) model, but that solutions with a range of $\nu$ and $\dot\nu$ 
are also statistically acceptable. To determine the search range of $\nu$ and 
$\dot\nu$ we found the 3$\sigma$ confidence ranges for each parameter 
from the frequency history fit. These considerations indicate that 
acceptable solutions are only possible within the following limits; 
$1.7553 \times 10^{-3} \; {\rm Hz} < \nu < 1.7578 \times 10^{-3} \; 
{\rm Hz}$, and $-1.7 \times 10^{-14} \; {\rm Hz \; s}^{-1} < \dot\nu < 9.0 
\times 10^{-15} \; {\rm Hz \; s}^{-1}$.

We next performed a grid search using all the data, sampling the range of 
$\nu$ and $\dot\nu$ found above from the time - frequency measurements. 
For each $\nu$ - $\dot\nu$ pair we calculated the total power $Z_2^2$ 
statistic as well as the $\chi^2$ statistic using the phase timing method. 
For the $\chi^2$ search we made phase measurements in all the good time 
intervals which were at least 400 s long. This resulted in a total of 
59 phase measurements. Our best solution has $\chi^2 = 62.5$ with 
$\nu = 1.7562465 \times 10^{-3}$ Hz, and $\dot\nu = 8 \times 10^{-18}$ Hz 
s$^{-1}$. Within the range of phase space searched there is only one other 
remotely plausible candidate solution. It has a $\chi^2 = 73.9$ with 
$\nu = 1.7570585 \times 10^{-3}$ Hz, and $\dot\nu = -1.0265 \times 10^{-14}$. 
The magnitude of the implied frequency derivative for this 
solution is orders of magnitude larger than expected for either gravitational 
radiation driven orbital decay or accretion-induced spin up of a white dwarf. 
These facts, combined with its substantially larger $\chi^2$, suggest that it 
can be excluded on astrophysical grounds and that it likely represents an 
``alias'' of the best solution produced by the sparse sampling. Although we
think this conclusion is fairly secure, it will require future timing 
measurements to test it definitively. 

Figure 3 shows contours of constant $\Delta\chi^2$ versus $\nu$ and $\dot\nu$ 
from our phase fitting analysis in the vicinity of our best solution. 
The results strongly favor a positive $\dot\nu = 8 \pm 3 \times 10^{-18}$ 
Hz s$^{-1}$. The quoted uncertainty on $\dot\nu$ is the 1$\sigma$, 
two-parameter confidence region. The results from the total $Z_2^2$ power 
method are entirely consistent with the phase fitting method, but in the 
interests of space we only show the $\Delta\chi^2$ contours. We obtained a 
minimum $\chi^2$ of 62.5, which for 57 degrees of freedom (dof), indicates 
that the data are consistent with the model. With $\dot\nu \equiv 0$ we have a
$\Delta\chi^2 = 11.5$, which excludes $\dot\nu = 0$ at the 99.7\% 
confidence level. Table 2 summarizes our best timing solution.

In Figure 4 we show the phase residuals from our best solution. 
The abscissa corresponds to phase measurement number (time ordered). 
The rms residual is $0.026$ cycles and is indicated by the dashed horizontal 
lines. Measurements from the 6 different observations (see Table 1) are 
denoted by the vertical dotted lines. We next used our best fitting $\chi^2$ 
model parameters to phase fold all the data. Figure 5 compares the phase
folded profiles from each individual observation. The average profile using 
all the data is also shown as the bottom trace. The profiles are more or less
consistent within the errors, thus it seems unlikely that pulse profile 
variations could have a significant influence on our timing analysis. 

\section{Implications and Discussion}

We have found strong evidence for an increase in the putative orbital 
frequency of RX J1914, however, processes other than gravitational radiation 
might also exert torques on the system that could produce orbital
period variations. In the remainder we discuss some of the issues regarding
interpretation of the observed period evolution as well as the implications
for the nature of RX J1914.

\subsection{Interpretation and Caveats}

Orbital period variations at the level of $\Delta\nu / \nu 
\lesssim 1\times 10^{-5}$ have been observed in a number of close binaries
(see Warner 1988; Applegate 1992; and references therein). These variations, 
typically observed over timescales of decades or longer, have been ascribed to 
magnetic activity of the low mass component (see Hall 1991; Applegate 1992; 
Arzoumanian, Fruchter \& Taylor 1994). In particular, the presence of 
conditions necessary to produce a stellar dynamo; convection and differential 
rotation, appear to be essential in producing such variations (Applegate 1992).
Applegate (1992) argues that in such systems the observed $\Delta\nu / \nu$ 
can be produced by a change in the quadrupole moment of the magnetically 
active star. If a change in the mass quadrupole is responsible for the 
observed orbital period variation in RX J1914, then it must have a magnitude 
$ ( \Delta q / m_2 R^2 ) = (\dot\nu \Delta T / 9 \nu_0) (a / R)^2 \approx 4 
\times 10^{-6}$, where for simplicity we have assumed the companion radius, 
$R$, is equal to the Roche lobe radius, and we have used the well known 
relation between orbital separation and Roche lobe radius (see for example 
Paczynski 1967). 

If RX J1914 is indeed a double-degenerate, then any type of standard magnetic
activity is not expected and it would seem unlikely that such an effect could 
be responsible for the observed $\dot\nu$.  However, non-degenerate, helium 
burning secondaries can develop surface convection zones (see Savonije et al. 
1986), which might arguably support a dynamo and drive a magnetic activity 
cycle. Orbital period changes due to magnetic activity are also necessarily 
accompanied by substantial optical luminosity variations (of order 0.1 mag). 
Although there does not appear to be any direct evidence for such optical 
variations, and we regard this scenario as unlikely, we cannot at present 
definitively rule it out. One way to do so will be to obtain long term X-ray 
monitoring of the orbital period, as well as further optical - infrared 
monitoring. Such observations will test our derived timing ephemeris and, if 
confirmed, will enable a secure association of the observed $\dot\nu$ with 
gravitational radiation. Additional clues would come from a constraint on 
$\ddot\nu$. If the only torque acting is gravitational radiation, then it 
should have a characteristic magnitude $\ddot\nu \approx 4 \times 10^{-31}$ Hz 
s$^{-2}$. If a $\ddot\nu$ term much greater than this is detected, then it 
will indicate the presence of additional torques in the system. 

\subsection{Orbital Evolution}

As mentioned above, the orbital evolution has important implications for the 
nature of mass transfer in close binaries. If one assumes that the system 
angular momentum is dominated by the orbital motion, and if one also 
imposes the assumptions of conservative mass exchange and angular momentum 
loss only from gravitational radiation, then it is relatively 
straightforward to show that the mass accretion rate, $\dot m_2$, and orbital 
frequency derivative, $\dot\nu$, are given by the following expressions 
(see for example, Rappaport, Joss \& Webbink 1982; Nelemans et al. 2001a);
\begin{equation}
\dot m_2 = - 1.72\times 10^{-7} \ \left ( \frac{m_2}{M_{\odot}} 
\right ) \left ( \frac{\mu}{M_{\odot}} \right ) \left ( \frac{a}{10^{10} \
{\rm cm}} \right )^2 \left ( \frac{\nu}{10^{-3} \ {\rm Hz}} \right )^4
\left ( \frac{1}{\left ( \frac{\xi (m_2)}{2} + \frac{5}{6} - q \right )} 
\right )\;\; M_{\odot} \; {\rm yr}^{-1} \; ,
\end{equation}
and
\begin{equation}
\dot\nu = -8.21 \times 10^{-18} \ \left ( \frac{\mu}{M_{\odot}} \right ) 
\left ( \frac{a}{10^{10} \ {\rm cm}} \right )^2 \left ( \frac{\nu}{10^{-3} \
{\rm Hz}} \right )^5 \left ( \frac{\left ( \frac{1}{3} - \xi (m_2) \right )}
{\left (\frac{\xi (m_2)}{2} + \frac{5}{6} - q \right )} \right )\; {\rm Hz} 
\ {\rm s}^{-1} \; .
\end{equation}
Here, $m_2$ is the mass of the donor star, $q \equiv m_2 / m_1$ is the mass
ratio ($q < 1$), and $\xi (m_2) = \frac{m_2}{r}\frac{d r}{d m_2}$, is the
dimensionless derivative of the radius of the donor, $r$, with respect to 
its mass. It can be shown that stable mass transfer requires, 
$\frac{\xi (m_2)}{2} + \frac{5}{6} > q$ (see, for example, Rappaport, Joss \& 
Webbink 1982). If the mass transfer is stable, the only way to have 
$\dot\nu > 0$, as observed, is for $\xi (m_2 ) > 1/3$. This argues against
degenerate donors since these stars grow as they lose mass and 
$\xi (m_2 ) < 0$. 

If the orbital decay results only from gravitational radiation losses and there
is no mass transfer, then the constraint on $\dot\nu$ implies a constraint on 
the so called ``chirp mass,''
\begin{equation} 
\left ( \frac{M_{ch}}{M_{\odot}} \right )^{5/3}  = 
\left ( \frac{\mu}{M_{\odot}} \right ) \ \left ( 
\frac{m_1 + m_2}{M_{\odot}} \right )^{2/3}  = 2.7 \times 10^{16} \left ( 
\frac{\nu}{10^{-3} \ {\rm Hz}} \right )^{-11/3} \ \dot\nu \; .
\end{equation}
This constraint follows directly from equation (6) and the use of Kepler's
law to substitute for the orbital separation, $a$. 
We show in Figure 6 the mass constraint derived from our $\dot\nu$ measurement.
The solid contour denotes the constraint for our best fit, while the 
dashed contours mark the $1\sigma$, two-parameter confidence limits. The 
inferred component masses are similar to those deduced by other researchers 
(see Ramsay et al. 2000; Wu et al. 2002; Marsh \& Steeghs 2002). 

\subsection{The Nature of the X-ray Flux}

Perhaps the simplest interpretation for the observed X-ray flux is that it is
accretion-driven (see, for example, Cropper et al. 1998; Ramsay et al. 2000). 
However, in the most straightforward accretion scenario, stable mass transfer 
would require that the donor be non-degenerate for reasons outlined above. 
Thus, our measurement of a positive $\dot\nu$ favors a non-degenerate donor
if accretion powers the X-ray flux. 

An alternative, proposed recently by Wu et al. (2002), is that the X-ray
flux is powered by a unipolar inductor mechanism similar to that which is 
thought to operate between Jupiter and Io (Clarke et al 1996). In this
model the donor sits inside its Roche lobe and no accretion takes place. 
An asynchronism between the orbital period and the spin of the secondary, 
on the order of a part in $10^{-3}$, would produce a voltage drop 
sufficient to account for the X-ray luminoisty.  
In this scenario, both electrical and gravitational dissipation will
cause the orbital frequency to increase, although gravitational losses 
dominate for most reasonable choices of the component masses (see Wu et al. 
2002). Thus, this model is consistent with our $\dot\nu$ measurement. A
criticism of the model is that the lifetime of the unipolar inductor 
phase should be relatively short as synchronization should occur in 
$\sim 1000$ yr (see Wu et al. 2002; Marsh \& Steeghs 2002), unless some 
additional torque acts to hold the system out of synchronization. An example
of a compact binary system which may be held out of synchronization is the 
``black widow'' pulsar PSR B1957+20. This object shows orbital period 
variations somewhat similar to those seen in magnetically active close
binaries (Arzoumanian, Fruchter \& Taylor 1994). Applegate \& Shaham (1994)
suggest that a wind driven from the companion by pulsar irradiation 
drives a torque which acts to keep the system asynchronized. This leads to
tidal dissipation which powers a wind from the companion and may also be 
responsible for the magnetic activity. Although this model does not appear 
directly applicable to RX J1914, it does suggest that other processes can 
maintain asynchronism over long timescales. 

Recently, Marsh \& Steeghs (2002) have proposed that RX J1914 is a 
``double-degenerate Algol,'' accreting by direct impact of the accretion
stream onto the primary. They argue that this model strongly favors a 
double-degenerate scenario for formation of AM CVn systems. In this model the
components are not magnetized and the spins are not necessarily synchronized, 
however, the X-ray source associated with impact of the stream is fixed in 
the rotating orbital reference frame. A positive $\dot\nu$ argues against this 
scenario for RX J1914 unless the donor is non-degenerate. A degenerate donor 
might still be viable if the mass transfer has gone unstable and formed a 
common envelope which is responsible for the observed orbital decay. However, 
this idea would appear to suffer from the same problem as the unipolar 
inductor model, that is, the lifetime of this phase would be short.  
Moreover, at some point the X-ray flux would become obscured by the envelope.  
If more complete theoretical investigations firmly rule out all non-degenerate 
donors because of the short orbital period then our positive $\dot\nu$ 
measurement would provide evidence for a unipolar induction mechanism for 
RX J1914.

Although the interpretation of the 569 s period as the orbital period seems 
most plausible, it is conceivable that the observed period is the spin 
period of an intermediate polar (IP), and that the orbital period has simply 
not been detected yet. Norton, Caswell \& Wynn (2002) have recently proposed
such a scenario for RX J1914. They suggest that the system is a stream-fed IP 
viewed nearly face-on.

If the system is an IP, then one would expect the 
orbital period to be longward of the 569 s spin period. Empirically, the spin 
and orbital periods of IPs satisfy $P_{orb} \approx 10 P_{spin}$, so that 
an orbital period of $\approx 100$ minutes would be reasonable. 
The mass donor in such a system would have a mass $\approx 0.1 M_{\odot}$,
implying an orbital separation, $a \approx 1 R_{\odot}$. Such an orbit would
introduce a phase delay to the white dwarf spin of at most a few seconds, 
which is much smaller than the phase offset implied for our best fit 
$\dot\nu$, so that the orbital modulation of a putative white dwarf spin 
would be negligible. If the primary white dwarf is accreting then it will 
experience a characteristic spin-up torque of 
magnitude,
\begin{equation}
N = \dot m \left ( G \ M \ r_m \right )^{1/2} \; ,
\end{equation}
where $r_m$ is the radius at which the accreting matter gets attached to the
magnetic field of the primary. This torque will produce a characteristic 
spin-up rate,
\begin{equation}
\dot\nu = 7.97 \times 10^{-17} \ \left ( \frac{\dot m}{10^{-8} \ M_{\odot} \
{\rm yr}^{-1}} \right ) \left ( \left ( \frac{M}{M_{\odot}} \right )^{-1} \ 
\left ( \frac{R}{10^{-2} \ R_{\odot}} \right )^{-3} \right )^{1/2} \; 
{\rm Hz} \ {\rm s}^{-1},
\end{equation}
where for simplicity we have assumed $r_m$ is equal to the white dwarf 
radius, $R$.
Interestingly, this $\dot\nu$ is not far from the observed value and can be 
made consistent with a reasonable choice of $\dot m$. Thus, the observed
$\dot\nu$ is not inconsistent with the IP model proposed by Norton, Caswell
\& Wynn (2002). Although orbital period variations in a truly face-on 
system would be very hard to detect, a further test of their hypothesis would 
be to probe deeper for an as yet unseen orbital period longward of the 569 s 
period. Such a search would ideally be made by Chandra 
or XMM-Newton, whose wide, eccentric orbits will provide better sensitivity 
than previous satellites hampered by the nature of the low Earth orbit 
observing window. 

\subsection{Gravitational Waves}

If the orbital period of RX J1914 is indeed 569 s, then its compact nature 
would make it a prime source of gravitational radiation for space-based 
detection. Its gravitational wave frequency $2\nu_{orb} = 3.51$ mHz is within 
the most sensitive frequency range of the proposed LISA interferometer 
(see Armstrong, Estabrook \& Tinto 2001), and is at high enough frequency to 
be outside the likely source confusion band produced by Galactic compact 
binaries (see Nelemans, Yungelson \& Portegies Zwart 2001b). The inferred 
strain amplitude for such systems, $\approx 1\times 10^{-21}$, is well above 
the predicted LISA noise floor at this frequency. Indeed, a detection of a
gravitational wave signal at twice the observed X-ray frequency would 
definitively rule out the IP models proposed for RX J1914.

If our timing solution is correct, then the phase advance due to $\dot\nu$ 
is approaching $0.3$ cycles, which can be easily confirmed with continued 
X-ray timing of the system, for example, with Chandra and XMM-Newton. If 
confirmed such observations will provide a much more accurate measurement 
of $\dot\nu$. Future monitoring will also establish if the observed $\dot\nu$ 
is related to the orbital period variations seen in magnetically active 
binaries. 

The combination of future X-ray timing and direct gravitational
wave measurements of systems like RX J1914 would provide a powerful new probe 
of gravitational radiation and binary evolution. The X-ray measurements will 
directly aid future detection of the system by space-based interferometers, 
such as LISA. Measurement of the orbital frequency derivative combined with 
detection from space of the time dependence of the gravitational wave amplitude
will provide the source distance (Schutz 1996). Moreover, the direct 
measurement of both orbital decay and the gravitational wave luminosity will 
provide new insights into the dynamics of orbital evolution in close binaries.

\subsection{Summary}

We have presented evidence that the 1.756 mHz X-ray frequency 
of RX J1914 is increasing at a rate of $8 \times 10^{-18}$ Hz s$^{-1}$. 
This rate is consistent with that expected from gravitational radiation 
losses in a detached compact binary and close to that expected given the 
likely scenarios which have been presented for the constituent masses
of RX J1914. If the X-ray emitting star were accreting from a degenerate donor,
then one would naively expect the orbit to be widening and the frequency 
decreasing, in conflict with the observations. This suggests that accretion 
may not power the X-ray flux, as proposed recently by Wu et al. (2002), who 
suggest a unipolar induction mechanism as the source of the X-rays. 
Alternatively, a non-degenerate helium burning secondary would be consistent 
with the observed $\dot\nu$; however, it may be that such systems cannot 
achieve such a short orbital period. Finally, an IP identification remains 
possible, if the system is nearly face-on. If this is the case, then the 
observed frequency increase likely reflects the accretion-induced spin up of 
the white dwarf primary. Although we favor the gravitational radiation orbit
decay interpretation, other compact, magnetically active binaries have shown
orbital period variations at a level similar to the implied orbital 
frequency change in RX J1914. It will have to wait for future monitoring of
the orbit to completely rule out this possibility.  

\acknowledgements

We thank Richard Mushotzky, Zaven Arzoumanian, Jean Swank and Craig Markwardt 
for many helpful comments and discussions. We thank the referee for a 
comprehensive review which helped us improve the paper. This work made use of 
data obtained from the High Energy Astrophysics Science Archive Research Center
(HEASARC) at Goddard Space Flight Center. 

\centerline{\bf References}

\noindent{} Armstrong, J. W., Estabrook, F. B. \& Tinto, M. 2001, Class. and
Quantum Gravity, 18, 4059

\noindent{} Arzoumanian, Z., Fruchter, A. S. \& Taylor, J. H. 1994, ApJ, 
426, L85

\noindent{} Applegate, J. H. 1992, ApJ, 385, 621

\noindent{} Applegate, J. H. \& Shaham, J. 1994, ApJ, 436, 312

\noindent{} Barrett, P.,  O'Donoghue, D., \&  Warner, B. 1988, MNRAS, 233, 759 

\noindent{} Buccheri, R. et al. 1983, A\&A, 128, 245

\noindent{} Clarke, J. T. et al. 1996, Science, 274, 404

\noindent{} Cropper, M. et al. 1998, MNRAS, 293, L57

\noindent{} Evans, C. R., Iben, I. \& Smarr, L. 1987, ApJ, 323, 129

\noindent{} Hall, D. S. 1991, ApJ, 380, L85

\noindent{} Hils, D. \& Bender, P. L.\ 2000, ApJ,, 537, 334

\noindent{} Hirayama, M. et al. 1996, ASCANews, \#4, ``Time Assignment of ASCA
GIS.'' http://heasarc.gsfc.nasa.gov/docs/asca/newsletters/Contents4.html

\noindent{} Iben, I. \& Tutukov, A. V. 1991, ApJ, 370, 615

\noindent{} Israel, G. L. et al. 1999, A\&A, 349, L1

\noindent{} Israel, G. L. et al. 2002, A\&A, submitted (astro-ph/0203043)

\noindent{} Marsh, T. R. \& Steeghs, D. 2002, MNRAS, in press, 
(astro-ph/0201309)

\noindent{} Motch, C. et al. 1996, A\&A, 307, 459

\noindent{} Nelemans, G., Portegies Zwart, S. F., Verbunt, F. \& Yungelson, 
L. R. 2001a, A\&A, 368, 939

\noindent{} Nelemans, G., Yungelson, L.~R., \& Portegies Zwart, S.~F.\ 2001b, 
A\&A, 375, 890

\noindent{} Norton, A. J., Haswell, C. A. \& Wynn, G. A. 2002, MNRAS, 
submitted, (astro-ph/0206013)

\noindent{} Paczynski, B. 1967, Acta Astron., 17, 287

\noindent{} Ramsay, G., Cropper, M., Wu, K., Mason, K.~O., \& Hakala, P.\ 
2000, MNRAS, 311, 75

\noindent{} Ramsay, G., et al. 2002, MNRAS in press, (astro-ph/0202281)

\noindent{} Ramsay, G. Hakala, P. \& Cropper, M. 2002, MNRAS, submitted 
(astro-ph/0203053)

\noindent{} Rappaport, S., Joss, P. C. \& Webbink, R. F. 1982, ApJ, 254, 616

\noindent{} Saito, Y. et al. 1997, ASCANews, \#5, ``Accuracy of the GIS Time
Assignment.'' 
http://heasarc.gsfc.nasa.gov/docs/asca/newsletters/Contents5.html

\noindent{} Savonije, G. J., de Kool, M. \& van den Heuvel, E. P. J. 1986, 
A\&A, 155, 51

\noindent{} Schutz, B. F. 1996, Class. and Quantum Gravity, 13, A219.

\noindent{} Standish, E. M., Newhall, X. X., Williams, J. G. \& Yeomans, D. K
            1992, in Explanatory Supplememt to the Astronomical Almanac, ed. 
            P. K. Seidelmann (Mill Valley: University Science), 239

\noindent{} Strohmayer, T. E., \& Markwardt, C. B., 2002, ApJ, in press, 
(astro-ph/0205435)

\noindent{} Strohmayer, T. E., \& Markwardt, C. B., 1999, ApJ, 516, L81

\noindent{} Taylor, J. H. \& Weisberg, J. M. 1989, ApJ, 345, 434

\noindent{} Tutukov, A., \& Yungelson, L. 1996, MNRAS, 280, 1035

\noindent{} Warner, B. 1995, {\it Cataclysmic Variable Stars}, Cambridge
Univ. Press, Cambridge UK.

\noindent{} Wu, K., Cropper, M., Ramsay, G. \& Sekiguchi, K. 2002, MNRAS 
in press, (astro-ph/0111358)

\pagebreak
\centerline{\bf Figure Captions}

\figcaption[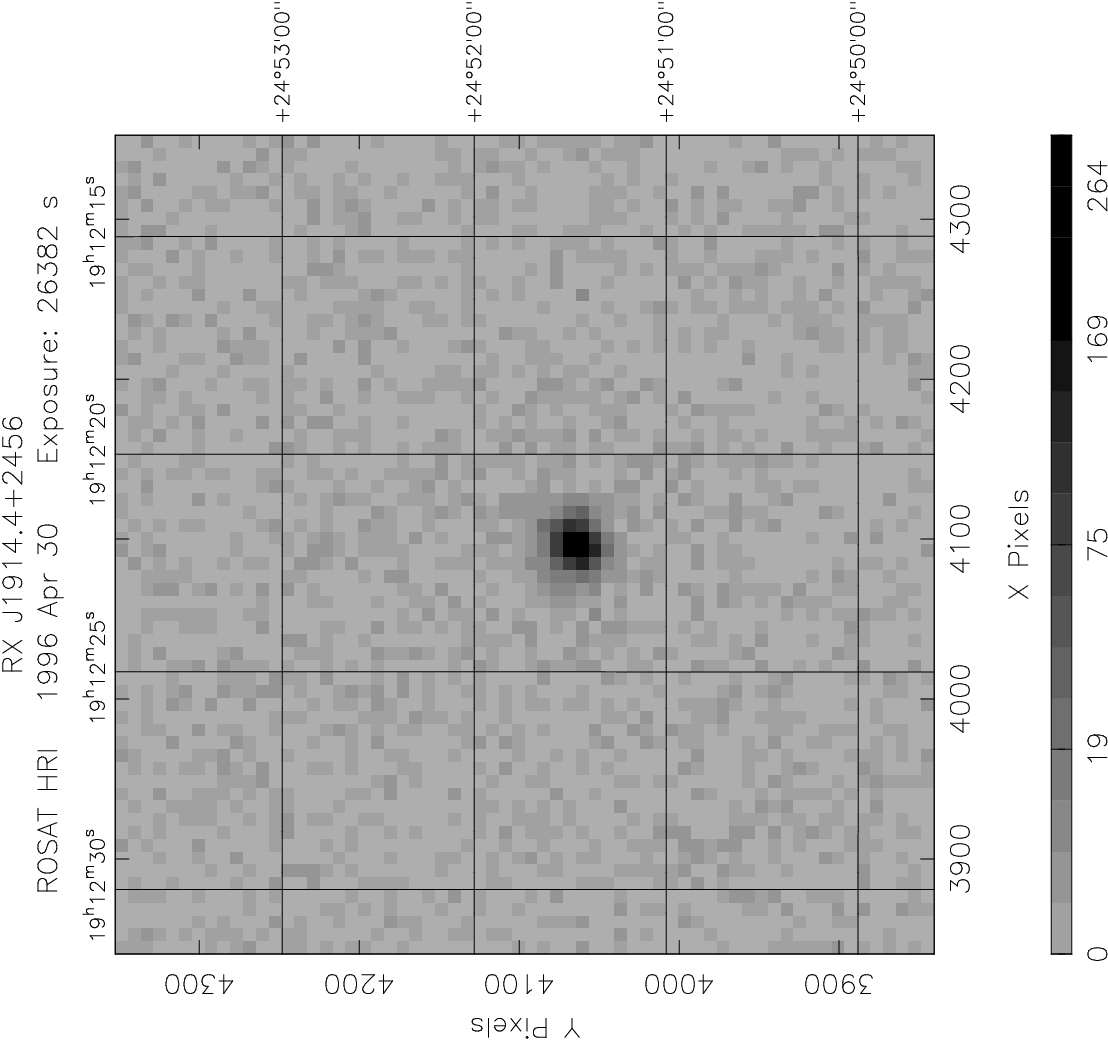]{A portion of the ROSAT HRI image of RX J1914.4+2456 from 
the 1996 April 30 observation. The grey-scale indicates counts in each pixel.
the pixels are 4'' on a side.  
\label{fig1}}

\figcaption[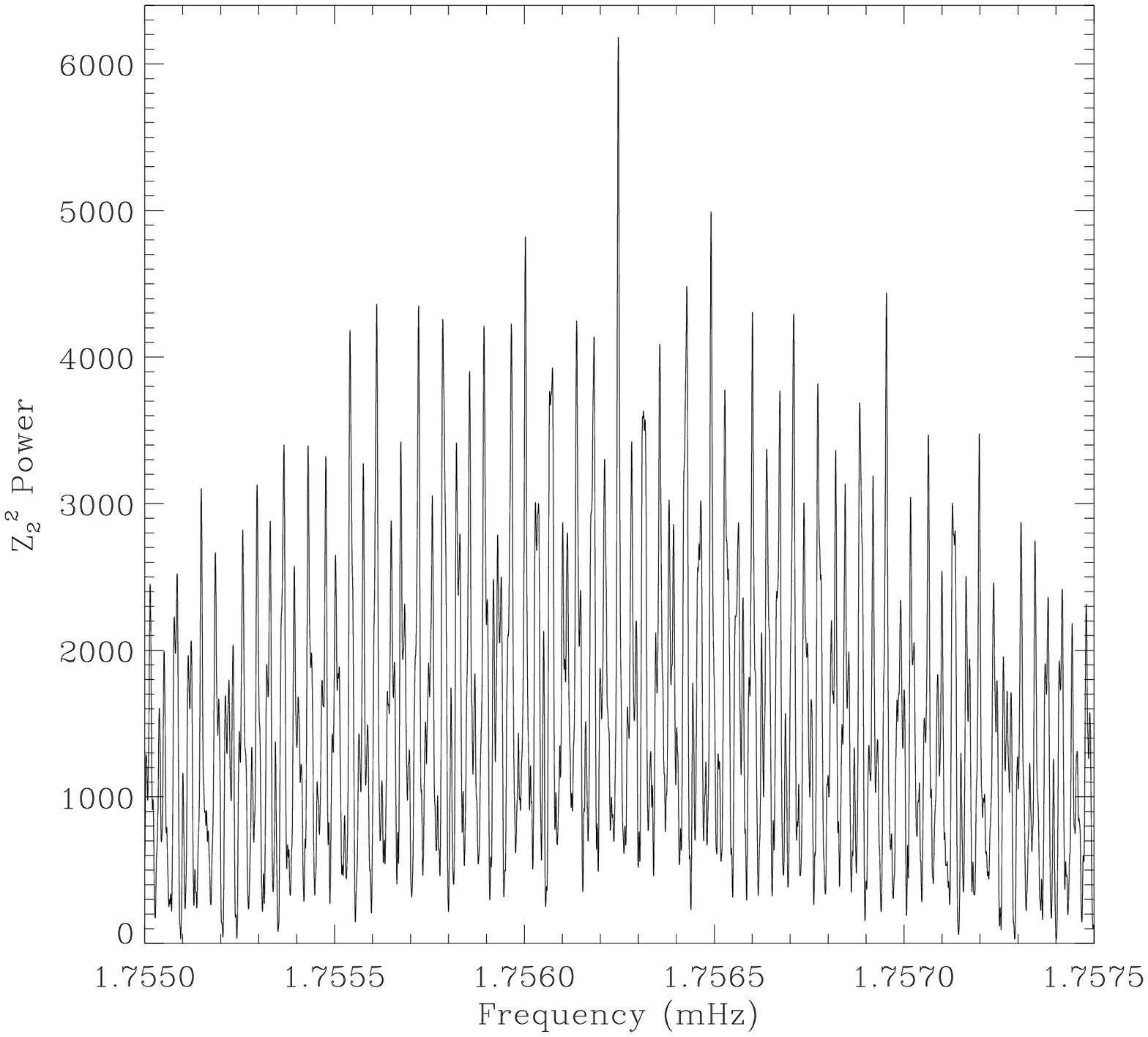]{Power density spectrum computed using the $Z_2^2$ power 
statistic, and $\dot\nu = 0$. The multiple side-lobe peaks result from the 
sparse sampling (i.e., temporal gaps) of the time series. However, as discussed
in detail in the text, the sampling is sufficient to resolve any cycle count 
ambiguities, that is, all the sub-peaks are significantly below the central 
peak.
\label{fig2}}

\figcaption[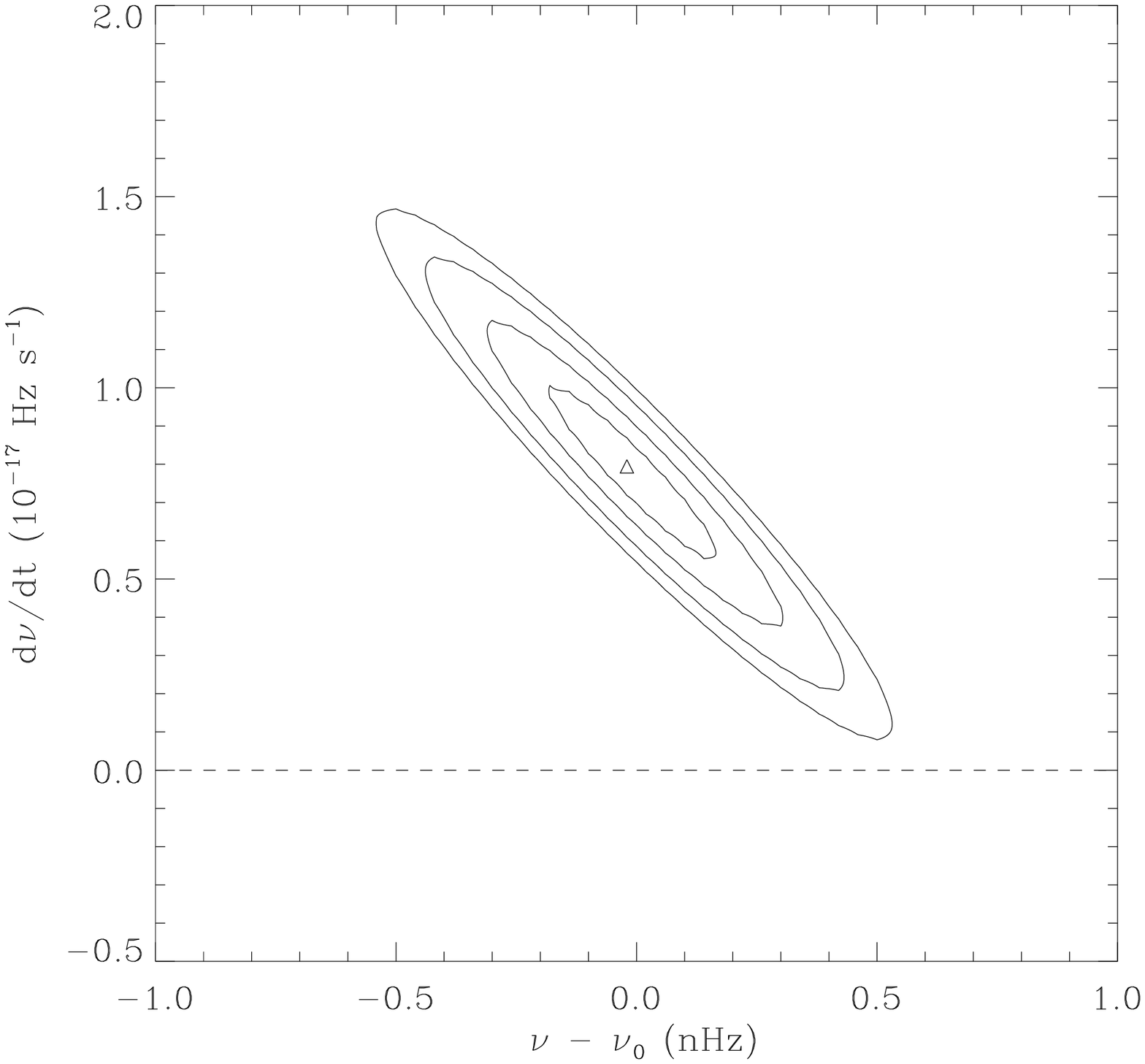]{Summary of our phase timing grid search for $\nu$ and 
$\dot\nu$. Shown is the map of constant $\Delta\chi^2 \equiv \chi^2 - 
min(\chi^2)$ in the vicinity of our best timing solution. Here, $\nu_0 
\equiv 1.7562465 \times 10^{-3}$ Hz. For $\Delta\chi^2$ we show contours at 
1, 3, 6, 9, and 12. The best fitting parameters are marked with the triangle 
symbol.
\label{fig3}}

\figcaption[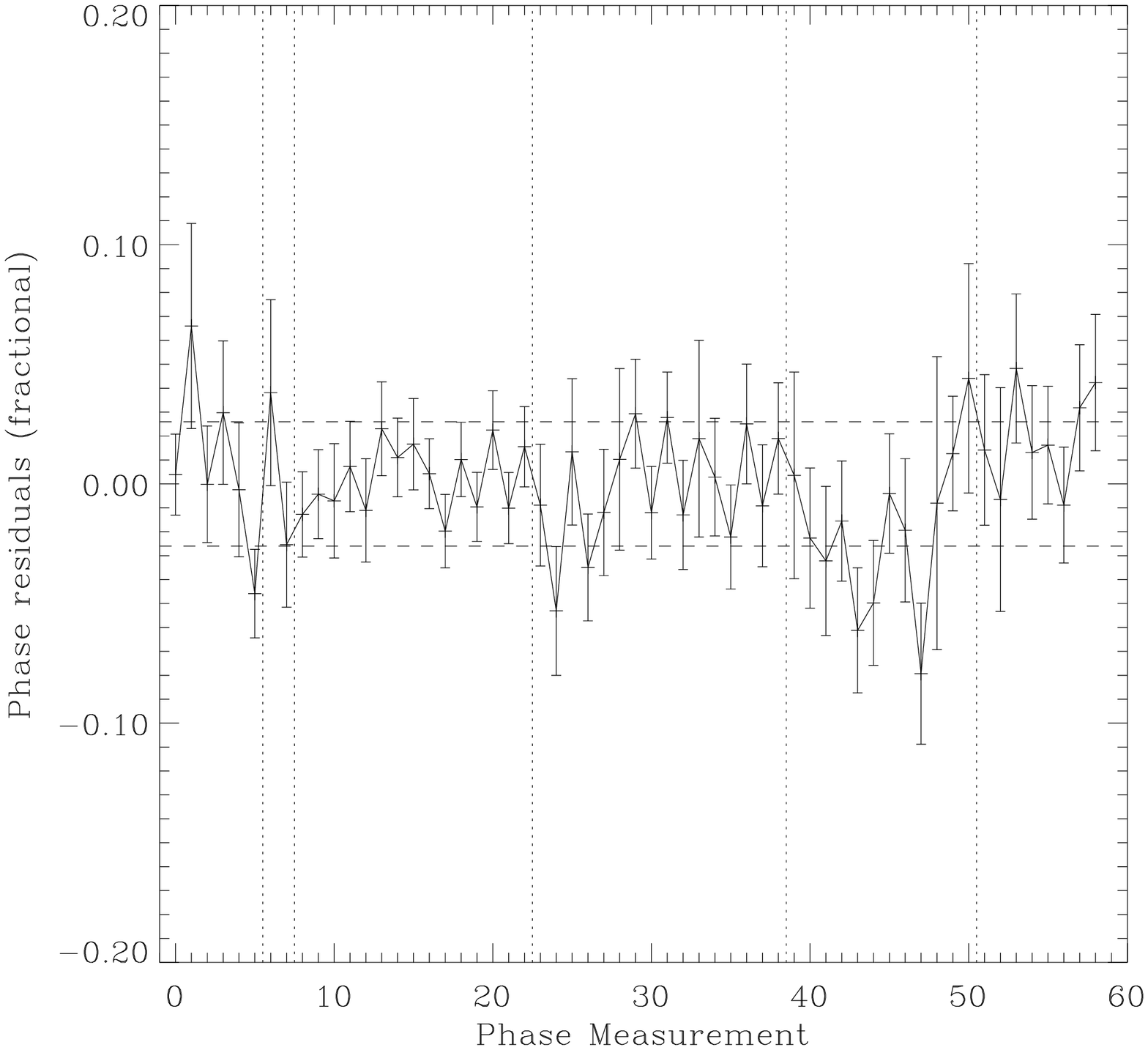]{Phase residuals (cycles) using the best fit parameters from 
minimizing $\chi^2$. The rms residual is $0.026$ cycles and is indicated by 
the dashed horizontal lines. The abscissa corresponds to phase measurement 
number (time ordered). Measurements from the 6 different observations 
(see Table 1) are denoted by the vertical dotted lines.
\label{fig4}}

\figcaption[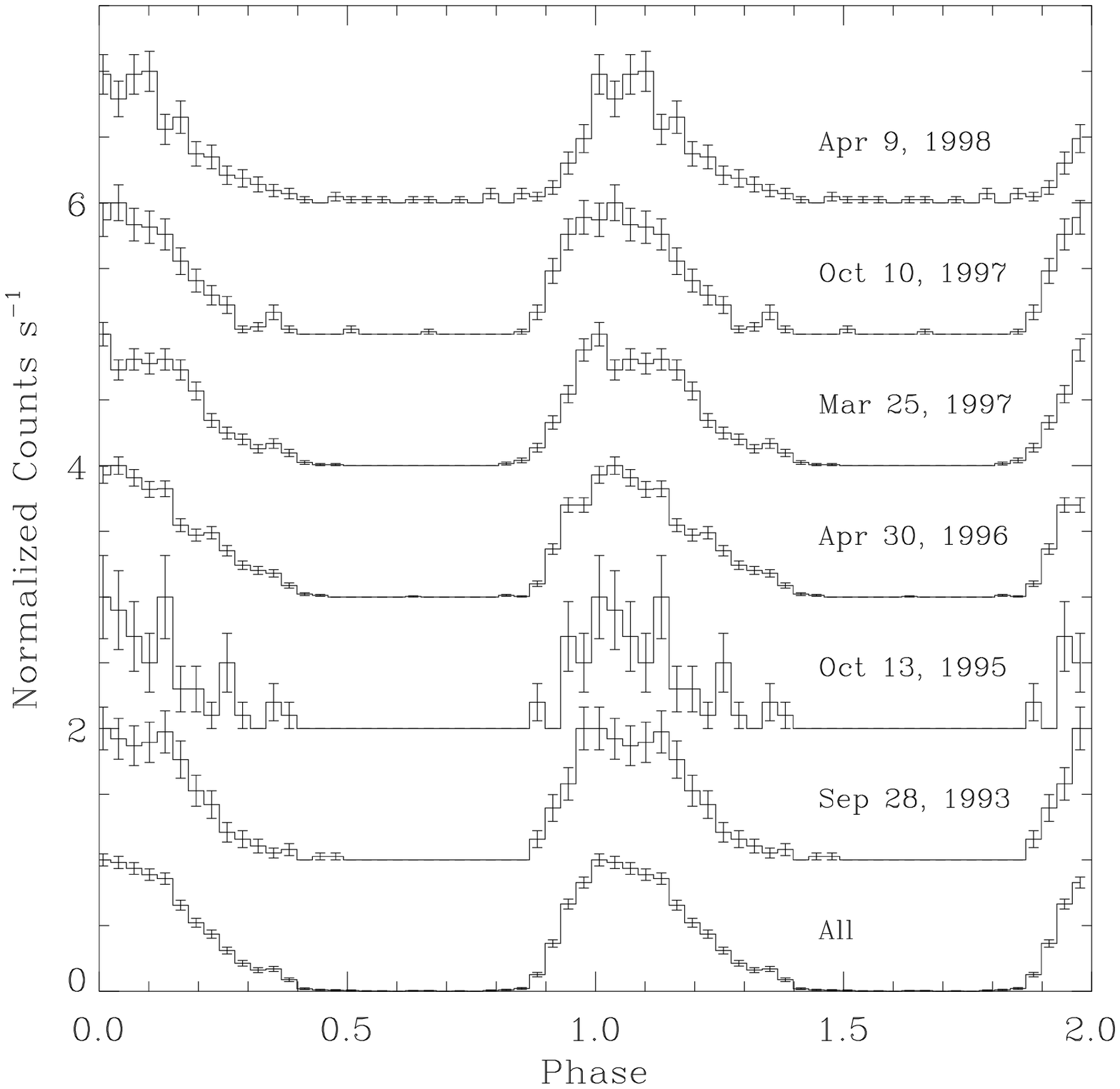]{Phase folded pulse profiles for each individual observation,
using our best $\chi^2$ solution. Each profile has been normalized to an 
amplitude of 1 and has been shifted vertically by 1 for clarity. Each 
observation is labelled with its start date (see also Table 1).
\label{fig6}}

\figcaption[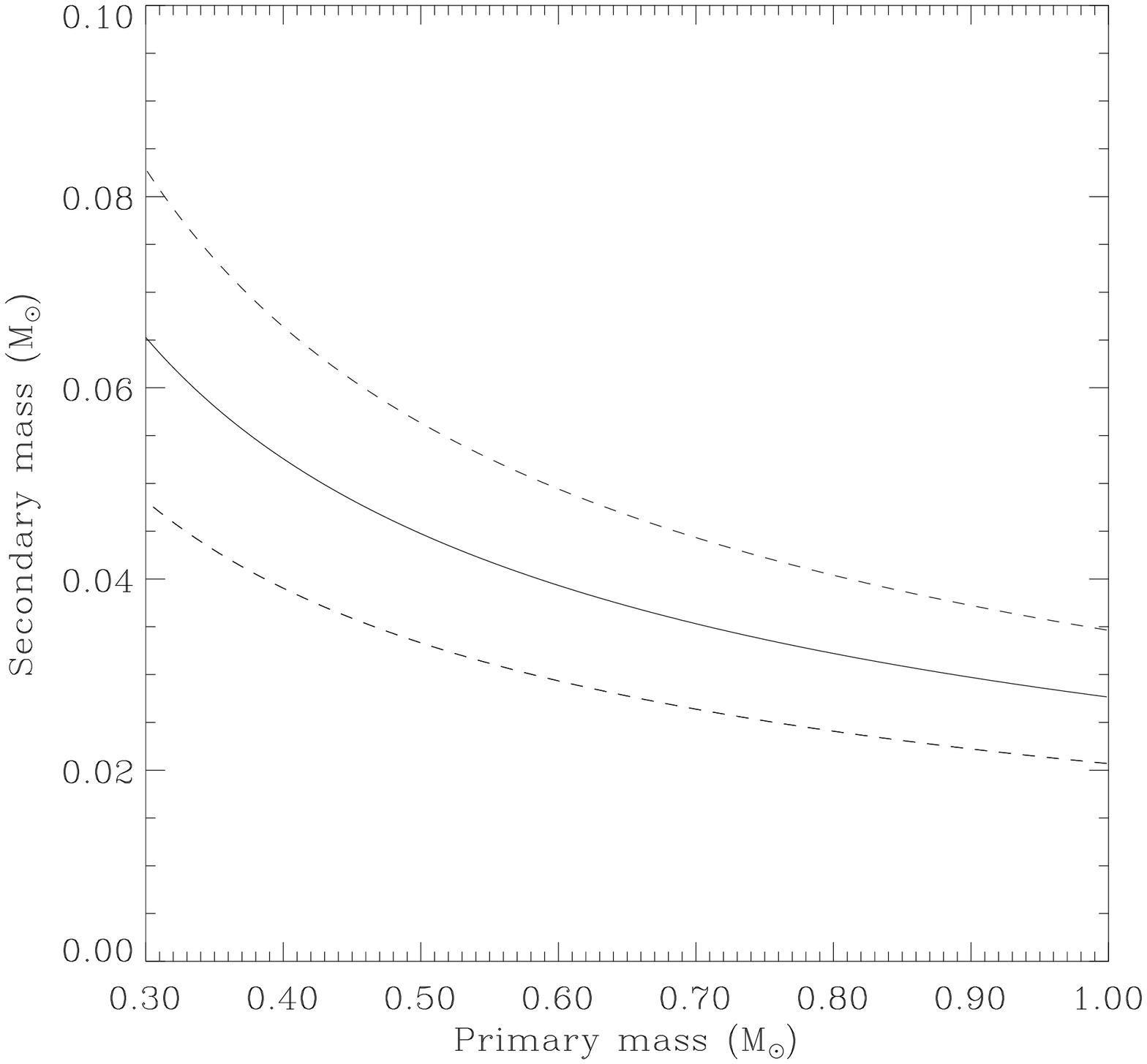]{Constraints on the component masses from our $\dot\nu$ 
measurement. The solid curve denotes the constraint for the best fitting 
$\dot\nu$, and the dashed contours denote the $1\sigma$ confidence interval. 
The constraint was derived assuming no mass transfer and gravitational 
radiation as the only angular momentum loss mechanism. 
\label{fig7}}

\pagebreak

\begin{figure}
\begin{center}
 \includegraphics[width=6in, height=6in,angle=-90]{f1.ps}
\end{center}
Figure 1: A portion of the ROSAT HRI image of RX J1914.4+2456 from 
the 1996 April 30 observation. The grey-scale indicates counts in each pixel.
the pixels are 4'' on a side.  
\end{figure}
\clearpage

\begin{figure}
\begin{center}
\includegraphics[width=6in, height=6in]{f2.ps}
\end{center}
Figure 2: Power density spectrum computed using the $Z_2^2$ power 
statistic, and $\dot\nu = 0$. The multiple side-lobe peaks result from the 
sparse sampling (i.e., temporal gaps) of the time series. However, as discussed
in detail in the text, the sampling is sufficient to resolve any cycle count 
ambiguities. That is, all the sub-peaks are significantly below the central 
peak.
\end{figure}
\clearpage

\begin{figure}
\begin{center}
 \includegraphics[width=6in, height=5in]{f3.ps}
\end{center}

Figure 3: Summary of our phase timing grid search for $\nu$ and 
$\dot\nu$. Shown is the map of constant $\Delta\chi^2 \equiv \chi^2 - 
min(\chi^2)$ in the vicinity of our best timing solution. Here, $\nu_0 
\equiv 1.7562465 \times 10^{-3}$ Hz. For $\Delta\chi^2$ we show contours at 
1, 3, 6, 9, and 12. The best fitting parameters are marked with the triangle 
symbol.
\end{figure}
\clearpage

\begin{figure}
\begin{center}
 \includegraphics[width=6in, height=6in]{f4.ps}
\end{center}
Figure 4: Phase residuals (cycles) using the best fit parameters from 
minimizing $\chi^2$. The rms residual is $0.026$ cycles and is indicated by 
the dashed horizontal lines. The abscissa corresponds to phase measurement 
number (time ordered). Measurements from the 6 different observations 
(see Table 1) are denoted by the vertical dotted lines.
\end{figure}
\clearpage

\begin{figure}
\begin{center}
 \includegraphics[width=6in, height=6in]{f5.ps}
\end{center}
Figure 5: Phase folded pulse profiles for each individual observation, using 
our best $\chi^2$ solution. Each profile has been normalized to an amplitude of
1 and has been shifted vertically by 1 for clarity. Each observation is 
labelled with its start date (see also Table 1).
\end{figure}
\clearpage
\clearpage

\begin{figure}
\begin{center}
 \includegraphics[width=6in, height=6in]{f6.ps}
\end{center}
Figure 6: Constraints on the component masses from our $\dot\nu$ measurement.
The solid curve denotes the constraint for the best fitting $\dot\nu$, and the
dashed contours denote the $1\sigma$ confidence interval. The constraint was
derived assuming no mass transfer and gravitational radiation as the only 
angular momentum loss mechanism. 
\end{figure}
\clearpage

\begin{table*}
\begin{center}{Table 1: ROSAT and ASCA Observations of RX J1914.4+2456}
\begin{tabular}{ccccc} \\
\tableline
\tableline
 OBSID     &  Instrument    &  Start UT    &  Stop UT  &  Exposure (ks) \\
\tableline
 300337  &  PSPC  & Sep 28, 1993 & Sep 29, 1993 & 7.0 \\
 300509  &  HRI   & Oct 13, 1995 & Oct 13, 1995 & 2.4 \\
 300509  &  HRI   & Apr 30, 1996 & May 5, 1996 & 25.9 \\
 300587  &  HRI   & Mar 25, 1997 & Mar 28, 1997 & 19.4 \\
 300337  &  HRI   & Oct 10, 1997 & Oct 11, 1997 & 21.7 \\
 36007000 & SIS  &  Apr 9, 1998 & Apr 10, 1998 &  21.4 \\
\tableline
\end{tabular}
\end{center}
\end{table*}

\begin{table*}
\begin{center}{Table 2: Timing Solution for RX J1914.4+2456}
\begin{tabular}{cc} \\
\tableline
\tableline
 Model Parameter &  Value \\
\tableline
 $t_0$ (TDB) & 49257.5333731 (MJD) \\
$\nu$ (Hz) & 0.0017562465(2) \\
$\dot\nu$ (Hz s$^{-1}$) & $0.8(3) \times 10^{-17}$ \\ 
\tableline
\end{tabular}
\end{center}
\end{table*}
\clearpage


\end{document}